\documentclass[aps,twocolumn]{revtex4}

\usepackage{graphics}
\usepackage{epsfig}
\usepackage{bm}
\usepackage{amsmath,amssymb}

\usepackage{graphics}
\usepackage{epsfig}

\usepackage{amsmath}
\usepackage{amssymb}
\usepackage{stmaryrd}

\def\ov#1{\overline{#1}}

\def\pd#1#2{\frac{\partial #1}{\partial #2}}

\def\cal#1{\mathcal{#1}}

\newcommand{\bc}{\begin{center}}
\newcommand{\ec}{\end{center}}
\newcommand{\bt}{\begin{tabbing}}
\newcommand{\et}{\end{tabbing}} 
\newcommand{\be}{\begin{eqnarray*}}
\newcommand{\ee}{\end{eqnarray*}}
\newcommand{\bs}{\begin{slide}}
\newcommand{\es}{\end{slide}}

\begin{document}

\title{Asymptotic Limit-cycle Analysis of the FitzHugh-Nagumo Equations}

\author{Alain J.~Brizard}
\affiliation{Department of Physics,  Saint Michael's College, Colchester, VT 05439, USA} 

\begin{abstract}
The asymptotic limit-cycle analysis of the FitzHugh-Nagumo equations is presented. In this work, we obtain an explicit analytical expression for the relaxation-oscillation period that is accurate within 1\% of their numerical values. In addition, we derive the critical parametric values leading to canard explosions and implosions in its associated limit cycles.
\end{abstract}

\date{\today}


\maketitle

\section{Introduction}

The FitzHugh-Nagumo (FHN) equations \cite{FitzHugh_1961,Nagumo_1962,FHN_Scholarpedia} provide a simple model describing the activation and deactivation of spiking behavior in neurons. Nagumo \cite{Nagumo_1962} introduced an electric-circuit representation of the FitzHugh \cite{FitzHugh_1961} model, in which a three-segment parallel circuit is built from a capacitor $C$ in one segment, in parallel with a tunnel diode (with an emf ${\cal E}_{0}$) in a second segment, and an $LR$-segment with a resistor $R$ connected in series with an inductor $L$.  

The Kirchhoff junction equation for the Nagumo circuit is expressed as the sum of three currents equal to the constant external current $I$ that flows into the three-segment junction:
\begin{equation}
I \;=\; I_{C} \;+\; I_{D}(\varepsilon) \;+\; I_{R}, 
\label{eq:I_K}
\end{equation}
where $I_{C}$ is the capacitor current, $ I_{D}(\varepsilon)$ is the diode current (which depends on the potential difference $\varepsilon$ across the diode), and $I_{R}$ is the current flowing through the $LR$ segment. 

By denoting the potential difference across each segment as $V$, we obtain the capacitor current $I_{C} = C\,dV/dt$, and the LR current $I_{R}$ yields the relation 
\begin{equation}
V \;=\; R\,I_{R} \;+\; L\,dI_{R}/dt. 
\label{eq:I_V}
\end{equation}
Lastly, we define the potential difference across the tunnel diode as $\varepsilon = V - {\cal E}_{0}$, so that the tunnel-diode current is modeled as
\begin{equation}
I_{D}(\varepsilon) \;=\; I_{0} \;-\; \frac{\Delta\varepsilon}{R_{0}} \left[ \left(\frac{\varepsilon - \varepsilon_{0}}{\Delta\varepsilon}\right) \;-\;  \frac{1}{3}  \left(\frac{\varepsilon - \varepsilon_{0}}{\Delta\varepsilon}\right)^{3} \right],
\label{eq:I_D}
\end{equation}
where $I_{0}$ flows through the diode when the potential difference is $\varepsilon_{0}$, which defines the negative resistance $1/I^{\prime}_{D}(\varepsilon_{0}) = -\,R_{0} < 0$. The potential differences $\varepsilon_{0} \pm \Delta\varepsilon$, on the other hand, are used to define the maximum and minimum $I_{D}(\varepsilon_{0} \mp \Delta\varepsilon) = I_{0} \pm \frac{2}{3}\,\Delta\varepsilon/R_{0}$ of the tunnel-diode current \eqref{eq:I_D}.

By introducing the following dimensionless variables: the diode potential $x = (\varepsilon - \varepsilon_{0})/\Delta\varepsilon$ and the resistor current $y = (I_{0} + I_{R})R_{0}/\Delta\varepsilon$, the Kirchhoff junction equation \eqref{eq:I_K} becomes
\begin{equation}
c \;=\; \dot{x} \;+\; x^{3}/3 \;-\; x \;+\; y,
\label{eq:x_FHN}
\end{equation}
where $c = IR_{0}/\Delta\varepsilon$ is the negative-resistance parameter, and we introduced the dimensionless time derivative $\dot{x} = R_{0}C\;dx/dt$, which is normalized to the $R_{0}C$ time constant. This equation is coupled to the resistor-current equation \eqref{eq:I_V}, now written in dimensionless form as
\begin{equation}
x \;=\; b\,y \;-\; a \;+\; \epsilon^{-1}\,\dot{y},
\label{eq:y_FHN}
\end{equation}
where $a = (RI_{0} + \varepsilon_{0} + {\cal E}_{0})/\Delta\varepsilon$ and $b = R/R_{0}$ are arbitrary constants, and the small dimensionless parameter is $\epsilon = \omega^{2}(R_{0}C)^{2} \ll 1$, where $\omega = 1/\sqrt{LC}$ is the natural $LC$ frequency (i.e., the LC period is chosen to be much longer than the $R_{0}C$ time constant).

There is a large amount of literature on the FHN equations and its extensions \cite{FHN_Scholarpedia}. As a simplification of the four-variable Hodgkin-Huxley model \cite{Hodgkin_Huxley_1952}, the FHN model \cite{Nagumo_1962} combines: (1) the membrane potential $V$ and the sodium activation variable $m$ as the membrane potential variable $x$; (2) the sodium inactivation variable $h$ and the potassium activation parameter $n$ as the recovery variable $y$; and (3) the membrane current is represented by the stimulus current $c$. Like the Van der Pol paradigm \cite{Kanamaru_2007}, these equations display a Hopf bifurcation at a critical value of the control parameter $c$, where a stable fixed point is replaced by a stable limit cycle. Once a stable limit cycle is created, a sudden transition from a small-amplitude oscillation to a large-amplitude relaxation oscillation is described as a canard explosion.

The remainder of the paper is organized as follows. In Sec.~\ref{sec:math}, we present the mathematical preliminary material that underlies the stability, bifurcation, and canard analysis of coupled first-order differential equations. In particular, we present the Fenichel geometric singular perturbation theory \cite{Fenichel_1979,Ginoux_2011}, which is applied to the Van der Pol equations. In Sec.~\ref{sec:FHN}, we apply this analysis to the FHN equations, which yields an explicit analytical expression for the relaxation-oscillation period that is accurate within 1\% of their numerical values, as well as critical parametric values leading to canard explosions and implosions in its associated limit cycles.

\section{\label{sec:math}Mathematical Preliminaries}

The FHN equations \eqref{eq:x_FHN}-\eqref{eq:y_FHN} are generically expressed as the nonlinear singular first-order ordinary differential equations
\begin{equation}
\left. \begin{array}{rcl}
\dot{x} &=& F(x,y; a) \\
\dot{y} &=& \epsilon\,G(x,y; a)
\end{array} \right\},
\label{eq:xy_general}
\end{equation}
where $x$ and $y$ denote dimensionless dynamical variables, and each dimensionless time derivative is represented with a dot (e.g., $\dot{x} = dx/dt$). On the right side of Eq.~\eqref{eq:xy_general}, the dimensionless parameter $\epsilon$ plays an important role in the qualitative solutions of Eq.~\eqref{eq:xy_general}, while the functions $F(x,y;a)$ and $G(x,y;a)$ (which may depend on a dimensionless control parameter $a$) are used to define the nullcline equations: $F(x,y;a) = 0 = G(x,y;a)$, which yield separate curves $y = f(x;a)$ and $y = g(x;a)$ onto the $(x,y)$-plane. A simplifying assumption used here is that the functions $F$ and $G$ are at most separately linear in $y$ and $a$, with $\partial^{2}F/\partial y\partial a = 0 = \partial^{2}G/\partial y\partial a$.

By introducing a new time normalization $t^{\prime} = \epsilon t$, the equations \eqref{eq:xy_general} may also be written as
\begin{equation}
\left. \begin{array}{rcl}
\epsilon\,x^{\prime} &=& F(x,y; a) \\
y^{\prime} &=& G(x,y; a)
\end{array} \right\},
\label{eq:xy_prime}
\end{equation}
where a prime now denotes a derivative with respect to $t^{\prime}$ (e.g., $x^{\prime} = dx/dt^{\prime}$). According to standard terminology, the times $t^{\prime}$ and $t$ are called the slow time and fast time, respectively, and Eqs.~\eqref{eq:xy_general} and
\eqref{eq:xy_prime} are called the fast system and slow system, respectively.

We note that the slope function $m(x,y;a) \equiv \dot{y}/\dot{x} = y^{\prime}/x^{\prime} = \epsilon\,G(x,y;a)/F(x,y;a)$ is a useful qualitative tool as we follow an orbit in the $y(t)$-versus-$x(t)$ phase space. In particular, we see that the orbit crosses the $y$-nullcline horizontally $(m = 0)$ while it crosses the $x$-nullcline vertically $(m = \pm\infty)$. Hence, in the limit $\epsilon \ll 1$, the slope function is near zero (i.e., the orbit is horizontal) unless the orbit is near the $x$-nullcline, where $F(x,y;a) \simeq 0$. As the slope $m(x,y;a)$ depends on the model parameter $a$, the shape of the orbit solution will also change with $a$.

The dynamical equations \eqref{eq:xy_general} can exhibit a type of large-amplitude oscillations called relaxation oscillations \cite{Strogatz_2015}. The paradigm for these large-amplitude oscillations is represented by the biased Van der Pol equation \cite{Diener_1984}
\begin{equation}
\frac{d^{2}x}{dt^{2}} \;-\; \nu\,\left(1 - x^{2}\right)\,\frac{dx}{dt} \;+\; \omega^{2}\,x \;=\; \omega^{2}\,a,
\label{eq:vdp_eq}
\end{equation}
where $\omega$ is the natural frequency of the linearized harmonic oscillator and $\nu$ is the negative dissipative rate, while the bias parameter $a$ represents an equilibrium value of the dimensionless oscillator displacement $x$. We note that the term $-\nu\,(1-x^{2})\,dx/dt$ yields negative dissipation in the range $x^{2} < 1$, which leads to exponential growth in that range.

From Eq.~\eqref{eq:vdp_eq}, we obtain the coupled dimensionless equations
\begin{equation}
\left. \begin{array}{rcl}
\dot{x} & = & x \;-\; x^{3}/3 \;-\; y \\
\dot{y} & = & \epsilon\,(x - a)
\end{array} \right\},
\label{eq:vdp_a}
\end{equation}
where the dimensionless time is normalized to $\nu^{-1}$ (i.e., $\dot{x} = \nu^{-1}dx/dt$) and $\epsilon \equiv \omega^{2}/\nu^{2}$ \cite{Footnote_VdP}. Here, the $x$-nullcline is $y(x) = x - x^{3}/3$ (which has a minimum at $x = -1$ and a maximum at $x = 1$) while the $y$-nullcline is a vertical line at $x = a$.

\begin{figure}
\epsfysize=3.3in
\epsfbox{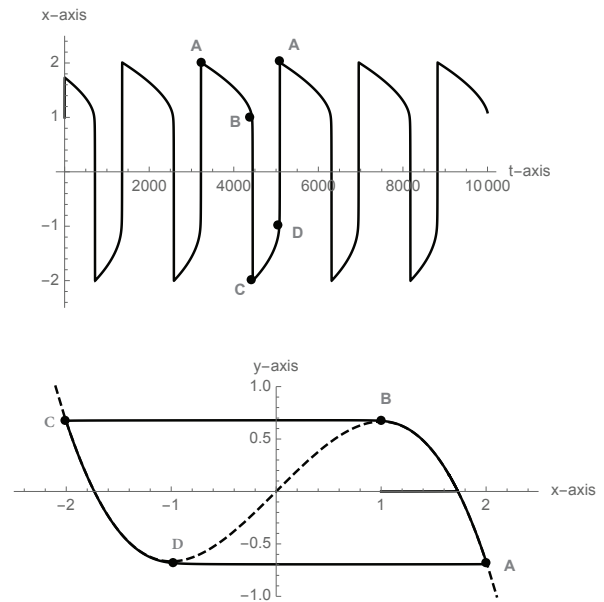}
\caption{Relaxation oscillation in the Van der Pol equations for $a = 0.5$ and $\epsilon = 0.001$. (Top) Plot of $x(t)$ versus time $t$ and (Bottom) Phase-space portrait showing $y(t)$ versus $x(t)$, with the $x$-nullcline shown as a dashed curve.}
\label{fig:VdP}
\end{figure}

\begin{figure}
\epsfysize=2in
\epsfbox{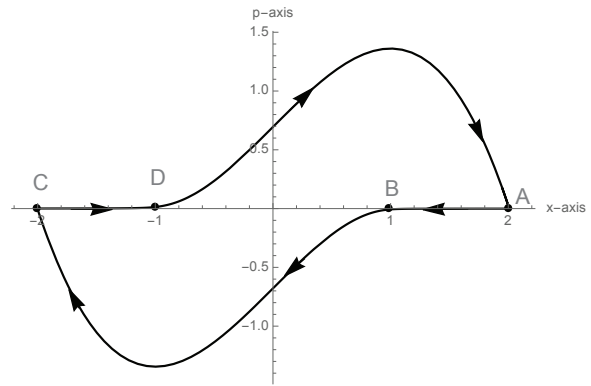}
\caption{Phase-space plot $p(t) \equiv \dot{x}(t) = x(t) - \frac{1}{3}\,x^{3}(t) - y(t)$ versus $x(t)$ for the relaxation oscillation shown in Fig.~\ref{fig:VdP}. Here, $y(t) \simeq x(t) - \frac{1}{3}\,x^{3}(t)$, i.e., the orbit is near the $x$-nullcline, when $p(t) \simeq 0$ from A to B and C to D.}
\label{fig:Phase_VdP}
\end{figure}

Figure \ref{fig:VdP} shows the solution of the Van der Pol equations \eqref{eq:vdp_a} for $a = 0.5$ and $\epsilon = 0.001$ and the initial conditions $x(0) = 1$ and $y(0) = 0$. In the top plot, the solution $x(t)$ shows slow orbits (on time scales of order $\epsilon^{-1}$) from A to B and C to D, and fast (exponential) transitions (on time scales of order $\epsilon^{\alpha}$, with $-1 < \alpha < 0$) from B to C and D to A. The bottom plot in Fig.~\ref{fig:VdP} shows that the slow orbits occur near the $x$-nullcline (shown as a dashed curve). Figure \ref{fig:Phase_VdP}, on the other hand, shows that the orbits from B to C and D to A include nonlinear exponential accelerations from $\pm 1$ to $\mp 1$, respectively, and then nonlinear exponential decays from $\mp 1$ to the turning points $x_{C} \simeq -2$ and $x_{A} \simeq 2$, respectively.

\subsection{Linear stability analysis}

If the nullcline curves of Eq.~\eqref{eq:xy_general} intersect at $(x_{0},y_{0})$, where $x_{0} = x_{0}(a)$ and $y_{0}(a) = f(x_{0}) = g(x_{0})$, the point $(x_{0},y_{0})$ is called a fixed point of Eq.~\eqref{eq:xy_general}. The stability of this fixed point is investigated through a standard normal-mode analysis \cite{Strogatz_2015}, where $x = x_{0} + \delta\ov{x}\,\exp(\lambda t)$ and $y = y_{0} + \delta\ov{y}\,\exp(\lambda t)$ are inserted into Eq.~\eqref{eq:xy_general} to obtain the linearized matrix equation
\begin{equation}
\left( \begin{array}{cc}
\lambda - F_{x0} & -\,F_{y0} \\
-\,\epsilon\,G_{x0} & \lambda - \epsilon\,G_{y0}
\end{array} \right) \cdot \left(\begin{array}{c} 
\delta\ov{x} \\
\delta\ov{y}
\end{array} \right) \;=\; 0,
\label{eq:Jac}
\end{equation}
where the constant eigenvector components $(\delta\ov{x},\delta\ov{y})$ are non-vanishing only if the determinant of the linearized matrix vanishes. Here, $(F_{x0},F_{y0})$ and $(G_{x0},G_{y0})$ are partial derivatives evaluated at the fixed point $(x_{0},y_{0})$ and the eigenvalues $\lambda_{\pm} = \frac{1}{2}\;\tau \pm \frac{1}{2}\, \sqrt{\tau^{2} - 4\;\Delta}$ are roots of the quadratic characteristic equation $\lambda^{2} - \tau\,\lambda + \Delta = 0$, where $\tau(a,\epsilon) \equiv F_{x0} + \epsilon\,G_{y0} = \lambda_{+} + \lambda_{-}$ and $\Delta(a,\epsilon) = \epsilon\,(F_{x0}\,G_{y0} - F_{y0}\,G_{x0}) = \lambda_{+}\cdot \lambda_{-}$ are the trace and determinant of the Jacobian matrix, respectively.

The fixed point is a stable point ($\tau < 0$ and $\Delta > 0$) that is either a node $(\tau^{2} > 4\;\Delta)$, when the eigenvalues are real and negative: $\lambda_{-} < \lambda_{+} < 0$, or a focus $(\tau^{2} < 4\;\Delta)$, when the eigenvalues are complex-valued ($\lambda_{-} = \lambda_{+}^{*}$) with a negative real part. Otherwise, the fixed point is either an unstable point ($\tau > 0$ and $\Delta > 0$) or a saddle point $(\Delta < 0)$. Periodic solutions of Eq.~\eqref{eq:xy_general} exist when a Hopf bifurcation \cite{Strogatz_2015} replaces an unstable fixed point with a stable limit cycle, which forms a closed curve in the $(x,y)$-plane. Here, a limit cycle appears when the $x$-nullcline function $f(x;a)$ has non-degenerate minimum and maximum points and it is stable whenever the trace $\tau(a) > 0$ is positive in the range $a_{s} < a < a_{u}$.

For the Van der Pol equations \eqref{eq:vdp_a}, we easily find the fixed point $(x_{0},y_{0}) = (a, a - a^{3}/3$) and the linear stability of that fixed point is described in terms of the trace $\tau = 1 - a^{2}$ and the determinant $\Delta = \epsilon > 0$. Here, the fixed point is stable if $a^{2} > 1$, and a limit cycle becomes stable in the range $-1 < a < 1$ as a result of a Hopf bifurcation \cite{Strogatz_2015} at $a = \pm 1$, where the fixed point merges with the critical points of the $x$-nullcline.

\subsection{Canard transition to relaxation oscillations}

Whenever the fixed point $x_{0}(a)$ of Eq.~\eqref{eq:xy_general} comes close to a critical point $x_{c}(a)$ of the $x$-nullcline, a sudden transition to a large-amplitude relaxation oscillation becomes possible. This transition, which occurs as the control parameter $a$ crosses a critical value $a_{c}(\epsilon)$, is referred to as a {\it canard} explosion or implosion, depending on whether the large-amplitude relaxation oscillation appears or disappears. For a brief review of the early literature on canard explosions, see Ref.~\cite{Diener_1984} and references therein. For a mathematical treatment, on the other hand, see Refs.~\cite{Krupa_2001,Fenichel_1979}.

We now present a perturbative calculation of the critical canard parameter $a_{c}(\epsilon)$ as an asymptotic expansion in terms of the small parameter $\epsilon$. For this purpose, we use the invariant-manifold solution $y = \Phi(x,\epsilon)$ of geometric singular perturbation theory \cite{Fenichel_1979,Ginoux_2011}, which yields the generic canard perturbation equation
\begin{eqnarray}
\dot{y} &=& \epsilon\,G\left(x, \Phi(x,\epsilon);\frac{}{}  a\right) \;=\; \pd{\Phi(x,\epsilon)}{x}\;\dot{x} \nonumber \\
 &=& \pd{\Phi(x,\epsilon)}{x}\;F\left(x, \Phi(x,\epsilon);\frac{}{} a\right),
 \label{eq:canard_eq}
\end{eqnarray}
where $\Phi(x,\epsilon) = \sum_{k=0}^{\infty}\epsilon^{k}\Phi_{k}(x)$ and $a_{c}(\epsilon) = \sum_{k=0}^{\infty}\epsilon^{k}a_{k}$. At the lowest order $(\epsilon = 0)$, we find
\begin{equation}
0 \;=\; F\left(x, \Phi_{0}(x);\frac{}{}a_{0}\right),
\label{eq:canard_eq_0}
\end{equation}
which yields the lowest-order $x$-nullcline 
\begin{equation}
\Phi_{0}(x) \;\equiv\; f(x;a_{0}). 
\end{equation}

\subsubsection{First-order perturbation analysis}

At the first order in $\epsilon$, we now find from Eq.~\eqref{eq:canard_eq}:
\begin{equation}
G(x,\Phi_{0};a_{0}) \;=\; \Phi_{0}^{\prime}(x) \left[ F_{y0}\;\Phi_{1}(x) \;+\frac{}{} F_{a0}\;a_{1} \right],
\label{eq:canard_1_eq}
\end{equation}
where $F_{y0} = (\partial F/\partial y)_{0} \neq 0$ and $F_{a0} = (\partial F/\partial a)_{0}$ are evaluated at $(x, \Phi_{0}; a_{0})$. Here, $\Phi_{0}^{\prime}(x)$ can be factored as
\begin{equation}
\Phi_{0}^{\prime}(x) \;\equiv\; (x - x_{c})\,\Psi_{0}(x),
\label{eq:Phi0_prime}
\end{equation}
where $\Psi_{0}(x)$ is assumed to be finite at the critical point $x = x_{c}(a_{0})$ (i.e., a minimum or a maximum of the $x$-nullcline). Since the right side vanishes at the critical point $x_{c}(a_{0})$, we find that $G(x_{c},\Phi_{0c};a_{0}) = 0$ implies the identity
\begin{equation}
x_{0}(a_{0}) \;\equiv\; x_{c}(a_{0}),
\label{eq:x_0c}
\end{equation}
where the fixed point $x_{0}$ has merged with the critical point $x_{c}$ of the $x$-nullcline at a unique value $a_{0}$, i.e., the fixed point $x_{0}(a_{0})$ is either at the maximum $x_{0}(a_{0}) = x_{B}(a_{0})$, which yields $a_{0} = a_{B0}$, or at the minimum $x_{0}(a_{0}) = x_{D}(a_{0})$, which yields $a_{0} = a_{D0}$. With this choice of $a_{0}$, we can write the factorization
\begin{equation}
G(x,\Phi_{0};a_{0}) \;\equiv\; (x - x_{c})\;H_{1}(x),
\label{eq:H1_factor}
\end{equation}
where $H_{1}(x)$ is finite at $x = x_{c}(a_{0})$. 

Hence, from Eq.~\eqref{eq:canard_1_eq}, we obtain the first-order solution
\begin{equation}
\Phi_{1}(x) \;\equiv\; K_{1}(x) \;-\; h(x)\;a_{1},
\label{eq:canard_1_sol}
\end{equation}
where we introduced the definitions
\begin{equation}
\left. \begin{array}{rcl}
K_{1}(x) &\equiv& H_{1}(x)/[\Psi_{0}(x)\,F_{y0}(x)] \\
 && \\
 h(x) &\equiv& F_{a0}(x)/F_{y0}(x)
 \end{array} \right\},
 \label{eq:canard_eq_1}
 \end{equation}
 which are both finite at $x = x_{c}(a_{0})$.

\subsubsection{Second-order perturbation analysis}

At the second order in $\epsilon$, we find from Eq.~\eqref{eq:canard_eq}:
\begin{eqnarray}
G_{y0}\;\Phi_{1} \;+\; G_{a0}\;a_{1} &=& \Phi_{0}^{\prime} \left( F_{y0}\;\Phi_{2} \;+\frac{}{} F_{a0}\;a_{2} \right) \nonumber \\
 &&+\; \Phi_{1}^{\prime} \left( F_{y0}\;\Phi_{1} \;+\frac{}{} F_{a0}\;a_{1} \right) \nonumber \\
  &=& \Phi_{0}^{\prime} F_{y0}\left( \Phi_{2} \;+\frac{}{} h\;a_{2} \right) \nonumber \\
 &&+\; F_{y0}\,\left( K_{1}^{\prime} \;-\frac{}{} h^{\prime}\,a_{1}\right) K_{1},
\label{eq:canard_eq_2}
 \end{eqnarray}
 where $G_{y0} = (\partial G/\partial y)_{0}$ and $G_{a0} = (\partial G/\partial a)_{0}$ are evaluated at $(x, \Phi_{0}; a_{0})$, and we have used the first-order solution \eqref{eq:canard_1_sol}. By rearranging terms in Eq.~\eqref{eq:canard_eq_2}, we obtain the second-order equation
\begin{equation}
S_{1}(x)\; a_{1} \;-\; R_{2}(x) \;=\; \Phi_{0}^{\prime}(x) \left[F_{y0}\;\Phi_{2}(x) \;+\frac{}{} F_{a0}\;a_{2} \right],
\label{eq:canard_2}
\end{equation}
where we introduced the definitions
\begin{eqnarray}
R_{2}(x) & = & K_{1}(x) \left[ F_{y0}\; K_{1}^{\prime}(x) \;-\frac{}{} G_{y0}\right], 
\label{eq:canard_R2} \\
S_{1}(x) & = &  G_{a0} -  G_{y0}\;h(x) + F_{y0}\,h^{\prime}(x)\,K_{1}(x),
\label{eq:canard_S1}
\end{eqnarray}
which are both finite at $x_{c}(a_{0})$.  

Once again, since the right side of this equation vanishes at the critical point $x = x_{c}(a_{0})$, the left side must also vanish at that point, and we obtain the first-order correction 
\begin{equation}
a_{1} \;=\; R_{2}(x_{c})/S_{1}(x_{c}).
\label{eq:canard_a1}
\end{equation}
By factoring the left side of Eq.~\eqref{eq:canard_2},
\begin{equation}
S_{1}(x)\; a_{1} \;-\; R_{2}(x) \;=\; (x - x_{c})\;H_{2}(x),
\label{eq:H2_factor}
\end{equation}
we now obtain the second-order solution
\begin{equation}
\Phi_{2}(x) \;\equiv\; K_{2}(x) \;-\; h(x)\,a_{2},
\label{eq:canard_2_sol}
\end{equation}
where $K_{2}(x) \equiv H_{2}(x)/[\Psi_{0}(x)F_{y0}(x)]$ and $h(x)$ is defined in Eq.~\eqref{eq:canard_eq_1}.

\subsubsection{Higher-order perturbation analysis}

By continuing the perturbation analysis at higher order $(n \geq 3)$, Eq.~\eqref{eq:canard_eq} yields the $n$th-order equation 
\begin{equation}
S_{1}(x)\,a_{n-1} - R_{n}(x) = \Phi_{0}^{\prime}(x)\,F_{y0}\,\left[\Phi_{n}(x) + h(x)\,a_{n}\right],
\label{eq:canard_n}
\end{equation}
where $S_{1}(x)$ is defined in Eq.~\eqref{eq:canard_S1} and
\begin{eqnarray}
R_{n}(x) & = & K_{1}(x)\,F_{y0}\,K_{n-1}^{\prime}(x) \;-\; G_{y0}\,K_{n-1}(x) \nonumber \\
 &&+ \sum_{k=1}^{n-2}F_{y0}\left[K_{k}^{\prime}(x) - h^{\prime}(x)\,a_{k}\right]K_{n-k}(x).
 \label{eq:Rn_sol}
 \end{eqnarray}
 Hence, the left side of Eq.~\eqref{eq:canard_n} vanishes at $x_{c}$ if 
 \begin{equation}
 a_{n-1} \;=\; R_{n}(x_{c})/S_{1}(x_{c}), 
 \label{eq:canard_an-1}
 \end{equation}
 and the $n$th-order solution is obtained by first obtaining the factorization 
 \begin{equation}
 S_{1}(x)\, a_{n-1} \;-\; R_{n}(x) \;=\; (x - x_{c})\,H_{n}(x), 
 \label{eq:Hn_factor}
 \end{equation}
 so that  
 \begin{equation}
\Phi_{n}(x) \;\equiv\; K_{n}(x) \;-\; h(x)\,a_{n},
\label{eq:canard_n_sol}
\end{equation}
where  $K_{n}(x) \equiv H_{n}(x)/[\Psi_{0}(x)F_{y0}(x)]$ and 
 \begin{equation}
 a_{n} \;=\; R_{n+1}(x_{c})/S_{1}(x_{c}), 
 \label{eq:canard_an}
 \end{equation} 
 is calculated from Eq.~\eqref{eq:Rn_sol}. We note that, once the function $R_{n}(x)$ is calculated in Eq.~\eqref{eq:Rn_sol}, the most computationally-intensive step is the factorization \eqref{eq:Hn_factor}, with $a_{n-1}$ calculated from Eq.~\eqref{eq:canard_an-1}.

As a result of the perturbative solution of Eq.~\eqref{eq:canard_eq}, we have, therefore, calculated the perturbation expansion of the canard critical parameter
\begin{equation}
a_{c}(\epsilon) \;=\; a_{0} \;+\; \frac{1}{S_{1}(x_{c})}\sum_{k=1}^{\infty}\epsilon^{k}\;R_{k+1}(x_{c}).
\label{eq:canard_01}
\end{equation}
For most applications, however, Eq.~\eqref{eq:canard_01} can be truncated at first order in the asymptotic limit $\epsilon \ll 1$: $a_{c}(\epsilon) \simeq a_{0} + a_{1}\,\epsilon$, where $a_{1} > 0$ for a canard explosion, while $a_{1} < 0$ for a canard implosion. 

\subsubsection{Van der Pol canard perturbation analysis}

\begin{figure}
\epsfysize=4in
\epsfbox{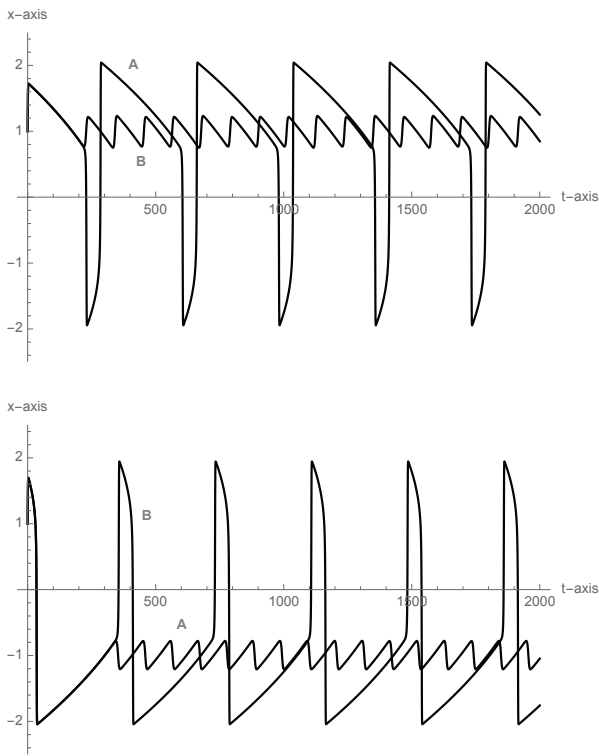}
\caption{Canard behavior in the Van der Pol equations for $\epsilon = 0.01$. (Top) Canard implosion when the large-amplitude relaxation oscillation suddenly disappears as $a = 0.998739\,(A) \rightarrow 0.998740\,(B)$. (Bottom) Canard explosion when the large-amplitude relaxation oscillation suddenly appears as $a = -0.998740\,(A) \rightarrow -0.998739\,(B)$.}
\label{fig:VdP_canard}
\end{figure}

Figure \ref{fig:VdP_canard} shows that the biased Van der Pol equations \eqref{eq:vdp_a} undergo canard explosion and implosion, when a small change in the bias parameter $a = -\,0.998740 \rightarrow -\,0.998739$ leads to the appearance of a large-amplitude relaxation oscillation from small-amplitude oscillations about the fixed point, while a small change in the bias parameter $a = 0.998739 \rightarrow 0.998740$ leading to the disappearance of large amplitude oscillations in $x(t)$ and $y(t)$ for the case $\epsilon = 0.01$. 

The canard perturbation equation \eqref{eq:canard_eq} for the Van der Pol equations \eqref{eq:vdp_a} is
\begin{equation}
\epsilon \left[ x \;-\frac{}{} a(\epsilon)\right] \;=\; \pd{\Phi(x,\epsilon)}{x} \left[ \Phi_{0}(x) \;-\frac{}{} \Phi(x,\epsilon)\right],
\end{equation}
where the partial derivatives evaluated at $\epsilon = 0$ are
\begin{equation}
\left. \begin{array}{rcl}
(F_{y0},\; F_{a0}) &=& (-1,\; 0) \\
 && \\
(G_{y0},\; G_{a0}) &=& (0,\; -1)
\end{array} \right\}.
\end{equation}
Here, the lowest-order solution $\Phi_{0}(x) = x - x^{3}/3$ has critical points at $x_{c} = \pm 1$ where $\Phi_{0}^{\prime}(x) = 1 - x^{2}$ vanishes. Hence, the lowest-order fixed point $x_{0} = a_{0}$ merges with the critical point $x_{c}$ when $a_{0} = \pm 1$. Because $F_{a0} = 0$, the function $h(x) = 0$ in Eq.~\eqref{eq:canard_eq_1}, while $\Psi_{0}(x) = x + a_{0}$ and $H_{1}(x) = -1$, so that $K_{1}(x) = 1/(x + a_{0}) = \Phi_{1}(x)$.

Next, in Eqs.~\eqref{eq:canard_R2}-\eqref{eq:canard_S1}, we have $R_{2} = -\,K_{1}\,K_{1}^{\prime} = 1/(x+a_{0})^{3}$ and $S_{1} = -1$, so that at $x = a_{0} = \pm 1$, we find the first-order correction $a_{1} = -1/(8\,a_{0}^{3})$, i.e., $a_{1} = -1/8$ for the canard implosion at $a_{0} = 1$, and $a_{1} = 1/8$ for the canard explosion at $a_{0} = -1$. 

For the canard explosion, the calculated critical parameter (truncated at first order) $a_{c}(\epsilon) = -1 + \epsilon/8$ yields $a_{c}(0.01) = -\,0.99875$, which is in excellent agreement with the numerical value $ -\,0.998740...$ shown in Fig.~\ref{fig:VdP_canard}. Because of the symmetry of the Van der Pol model, the calculated critical parameter (truncated at first order) $a_{c}(\epsilon) = 1 - \epsilon/8$ for the canard implosion yields $a_{c}(0.01) = 0.99875$, which is again in excellent agreement with the numerical value $0.998740...$ shown in Fig.~\ref{fig:VdP_canard}. Higher-order corrections to the Van der Pol canard parameter $a_{c}(\epsilon) = 1 - \epsilon/8 - 3\,\epsilon^{2}/32 - 173\,\epsilon^{3}/1024 - \cdots$ can be computed up to arbitrary order \cite{Algaba_2020} but they are not needed in what follows.

\subsection{Asymptotic limit-cycle period}

We saw in Figs.~\ref{fig:VdP}-\ref{fig:Phase_VdP} that, in the asymptotic limit $\epsilon \ll 1$, the limit-cycle curve of Eq.~\eqref{eq:xy_general} is composed of slow segments that are close to the $x$-nullcline. In this limit, the asymptotic period can be calculated as follows. First, we begin with the $x$-nullcline $y = f(x;a)$ on which we obtain $dy/dt = f^{\prime}(x;a)\,dx/dt$. Next, we use the $y$-equation $dy/dt = \epsilon\,G(x,y;a)$, into which we substitute the $x$-nullcline equation: $dy/dt = \epsilon\,G(x,\,f(x;a);a)$. 

By combining these equations, we obtain the infinitesimal asymptotic-period equation 
\[ \epsilon\,dt \;=\; f^{\prime}(x;a)\,dx/G\left(x, f(x;a);a\right), \]
which yields the asymptotic limit-cycle period
\begin{eqnarray}
\epsilon\,T_{\rm ABCDA}(a) &=& \int_{x_{A}(a)}^{x_{B}(a)} \frac{f^{\prime}(x;a)\;dx}{G(x,f(x;a);a)} \nonumber \\
 &&+\; \int_{x_{C}(a)}^{x_{D}(a)} \frac{f^{\prime}(x;a)\;dx}{G(x,f(x;a);a)}.
\label{eq:vdP_period}
\end{eqnarray}
Here, the asymptotic limit cycle ABCDA combines the slow $x$-nullcline orbits $x_{A} \rightarrow x_{B}$ and $x_{C} \rightarrow x_{D}$ and the fast horizontal transitions $x_{B} \rightarrow x_{C}$ and $x_{D} \rightarrow x_{A}$, which are ignored in Eq.~\eqref{eq:vdP_period}. Generically, the values $x_{D}(a) < x_{B}(a)$ are the minimum and maximum of the $x$-nullcline $y = f(x;a)$, respectively, where $f^{\prime}(x;a)$ vanishes. The points $x_{C}(a) < x_{A}(a)$, on the other hand, are the minimum and maximum of the asymptotic limit cycle.

\begin{figure}
\epsfysize=1.8in
\epsfbox{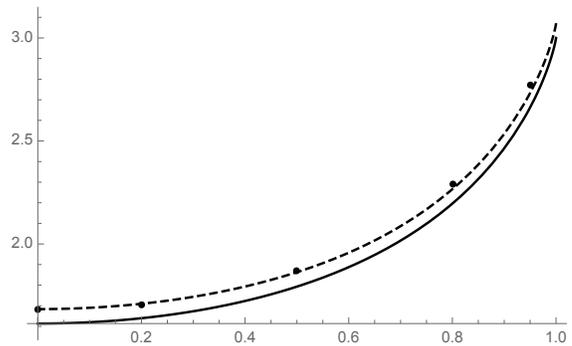}
\caption{Plots of the asymptotic Van der Pol period $\epsilon\,T_{\rm VdP}(a)$ (solid) and the corrected asymptotic Van der Pol period $\epsilon\,T^{\alpha}_{\rm VdP}(a,\epsilon)$ (dashed) versus the bias parameter $a$, in the limit $\epsilon = 0.001 \ll 1$. The numerical periods $\epsilon\,T_{\rm num}(a,\epsilon)$, shown as dots, are approximately 4\% higher than the asymptotic Van der Pol period \eqref{eq:VdP_period_a} and are within 1\% of the corrected asymptotic Van der Pol period \eqref{eq:VdP_period_alpha} .}
\label{fig:VdP_period}
\end{figure}

In the limit $\epsilon \ll 1$,  the phase-space portrait for the Van der Pol equations \eqref{eq:vdp_a} shown in Fig.~\ref{fig:VdP} has slow segments $A\,(x_{A} = 2) \rightarrow B\,(x_{B} = 1)$ and $C\,(x_{C} = -2) \rightarrow D\,(x_{D} = -1)$ on the $x$-nullcline (shown as a dashed curve) and fast horizontal transitions $B\,(x_{B} = 1) \rightarrow C\,(x_{C} = -2)$ and $D\,(x_{D} = -1) \rightarrow A\,(x_{A} = 2)$. The asymptotic period \eqref{eq:vdP_period} for the Van der Pol limit-cycle ABCDA is calculated as
\begin{eqnarray}
\epsilon\,T_{\rm VdP}(a) &=& \int_{2}^{1}\frac{(1 - x^{2})\,dx}{x - a} \;+\; \int_{-2}^{-1}\frac{(1 - x^{2})\,dx}{x - a} \nonumber \\
 &=& 3 \;-\; (1 - a^{2})\;\ln\left(\frac{4 - a^{2}}{1 - a^{2}}\right),
\label{eq:VdP_period_a}
\end{eqnarray}
which is shown in Fig.~\ref{fig:VdP_period} as a solid curve. We note that the asymptotic Van der Pol period \eqref{eq:VdP_period_a} is symmetric in $a$, i.e., $T_{\rm VdP}(-a) = T_{\rm VdP}(a)$. 

The next term in the asymptotic expansion of the Van der Pol period \eqref{eq:VdP_period_a} involves a nontrivial correction associated with the complex orbits seen in Fig.~\ref{fig:Phase_VdP} on their way to the turning points at $x_{A,C} \simeq \pm 2$ \cite{Strogatz_2015}. This correction is expressed as $3\,\alpha\,\epsilon^{2/3}$ \cite{Bender_Orszag_1978}, where $\alpha = 2.338107...$ denotes the lowest zero of the Airy function ${\rm Ai}(-x)$. If we add this correction to the Van der Pol asymptotic period \eqref{eq:VdP_period_a}, we obtain
\begin{equation}
\epsilon\,T^{\alpha}_{\rm VdP}(a,\epsilon) \;\equiv\; 3 \;-\; (1 - a^{2})\;\ln\left(\frac{4 - a^{2}}{1 - a^{2}}\right) \;+\; 3\,\alpha\,\epsilon^{2/3},
\label{eq:VdP_period_alpha}
\end{equation}
where the correction is assumed to be independent of the bias parameter $a$ (a more thorough calculation, which is omitted here, would be required to explore this dependence).

The numerical periods $\epsilon\,T_{\rm num}(a,\epsilon)$, which are shown in Fig.~\ref{fig:VdP_period} as dots, are within 4\% higher than the asymptotic Van der Pol period \eqref{eq:VdP_period_a} and are within 1\% of the corrected asymptotic Van der Pol period \eqref{eq:VdP_period_alpha}. These numerical results show that the asymptotic limit $\epsilon \ll 1$ enables us to evaluate the limit-cycle period according to Eq.~\eqref{eq:VdP_period_alpha} with excellent accuracy, on both qualitative and quantitative basis.

\section{\label{sec:FHN}FitzHugh-Nagumo Equations}

The FHN equations \cite{FHN_Scholarpedia} offer a simple model used to study the conditions leading to firing of neuron cells. Here, the FHN equations are expressed as
\begin{eqnarray}
\dot{x} & = & x \;-\; x^{3}/3 \;+\; c \;-\; y, \label{eq:x_dot} \\
\dot{y} &=& \epsilon \left( x \;+\frac{}{} a \;-\; b\,y\right), \label{eq:y_dot}
\end{eqnarray}
where $(a,b,c)$ are constants and $\epsilon \ll 1$. In what follows, we will use the model parameters $(a, b) = (3/5, 4/5)$ for the purpose of explicit calculations and numerical simulations, and the control parameter $c$ will determine the type of solutions for Eqs.~\eqref{eq:x_dot}-\eqref{eq:y_dot}.

\begin{figure}
\epsfysize=1.8in
\epsfbox{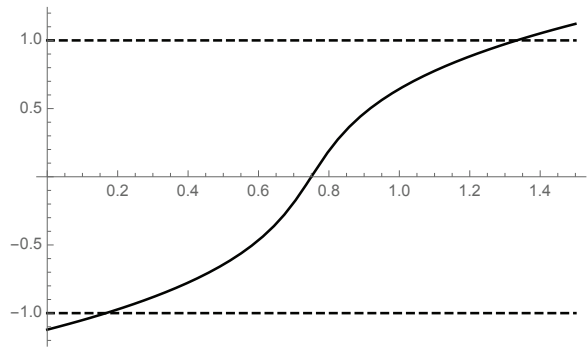}
\caption{Plot of the fixed point $x_{0}(c) = \sinh[\psi(c)/3]$ as a function of the control parameter $c$. The fixed point reaches the critical points $\pm 1$ (dashed lines) of the $x$-nullcline at $c = 1/6$ and $c = 4/3$.}
\label{fig:FHN_fix}
\end{figure}

The FHN nullcline equations are
\begin{equation}
\left. \begin{array}{rl}
x-{\rm nullcline}: & f(x) = x \;-\; x^{3}/3 \;+\; c \\
 & \\
 y-{\rm nullcline}: & g(x) = (5\,x + 3)/4 
 \end{array} \right\},
 \end{equation}
which intersect at a single fixed point $(x_{0}, y_{0})$, where $x_{0}(c)$ is the single real root of the cubic equation
 \begin{equation}
 4\,x^{3} \;+\; 3\,x \;-\; \left(12\,c \;-\; 9 \right) \;=\; 0.
 \label{eq:FHN_cubic}
 \end{equation}
The three roots of this equation \cite{Brizard_2014} are
\begin{eqnarray}
x_{1}(c) & = & i\cos\left(\frac{\pi}{6} - i\,\frac{\psi(c)}{3}\right), \label{eq:x1_c} \\
x_{2}(c) & = & -\,i\cos\left(\frac{\pi}{2} - i\,\frac{\psi(c)}{3}\right) = \sinh\left(\frac{1}{3}\,\psi(c)\right),  \label{eq:x2_c}\\
x_{3}(c) & = & -\,i\cos\left(\frac{\pi}{6} + i\,\frac{\psi(c)}{3}\right) \equiv x_{1}^{*}(c), \label{eq:x3_c}
\end{eqnarray}
where $\psi(c) \equiv {\rm arcsinh}(12\,c - 9)$. Here, the fixed point $x_{0}(c) = x_{2}(c) = \sinh[\psi(c)/3]$ reaches the critical points $\pm 1$ of the $x$-nullcline at $c = 1/6$ and $c = 4/3$, respectively (see Fig.~\ref{fig:FHN_fix}).

\subsection{Linear stability of the fixed point}

\begin{figure}
\epsfysize=1.8in
\epsfbox{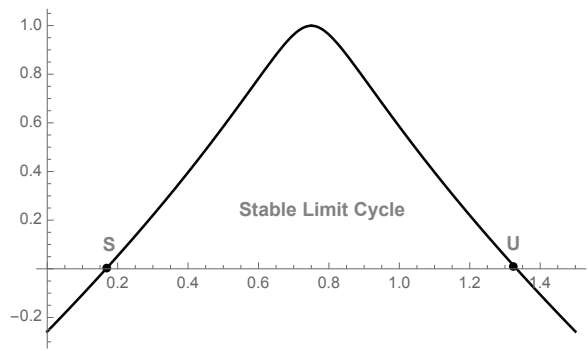}
\caption{Linear stability diagram for the FHN equations for $\epsilon = 0.001$. Trace $\tau(c)$ versus $c$, showing a stable limit cycle in the range $c_{s}(\epsilon) < c < c_{u}(\epsilon)$.}
\label{fig:FHN_stability}
\end{figure}

The linear stability of the fixed point $(x_{0},y_{0})$ is determined from the Jacobian matrix
\begin{equation}
{\sf J}_{0}(c,\epsilon) \;=\; \left( \begin{array}{cc} 
1 \;-\; x_{0}^{2}(c) & -\,1 \\
\epsilon & -\,4\,\epsilon/5
\end{array} \right),
\end{equation}
where the trace is $\tau = (1 - 4\,\epsilon/5) - x_{0}^{2}$ and the determinant is $\Delta = \epsilon\,(1 + 4\,x_{0}^{2})/5 > 0$. Marginal stability $(\tau = 0)$ occurs at $c_{s}(\epsilon) = 3/4 - \delta(\epsilon)/12$ and $c_{u}(\epsilon) = 3/4 + \delta(\epsilon)/12$, where $\delta(\epsilon) = (7 - 16\,\epsilon/5)\sqrt{1 - 4\epsilon/5} < 7$ for $\epsilon > 0$. Here, the fixed point is stable if $c < c_{s}(\epsilon)$ and $c > c_{u}(\epsilon)$, while a limit cycle is stable (for $\epsilon = 0.001$) in the range
\begin{equation}
 c_{s}(\epsilon) \;=\; 0.167167 \;<\; c \;<\; c_{u}(\epsilon) \;=\; 1.33283.
 \end{equation}
Here, we note that $c_{s}(\epsilon) > 3/4 - 7/12 = 1/6$ and $c_{u}(\epsilon) < 3/4 + 7/12 = 4/3$, i.e., the fixed point loses stability after it has reached the $x$-nullcline minimum at $x = -1$, while it regains stability before it has reached the $x$-nullcline maximum at $x = 1$.

\begin{figure}
\epsfysize=2.8in
\epsfbox{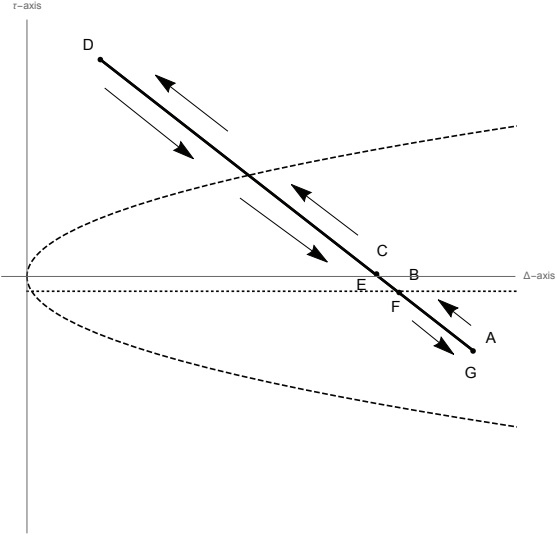}
\caption{Path in the stability (trace-versus-determinant) space for $0 \leq c \leq 3/2$. The dashed parabola denotes $\Delta(c) = \tau^{2}(c)/4$ and the horizontal dotted line at $\tau(c) = -\,4\epsilon/5$ represents the location where $x_{0}(c) = \pm1$. The description of the path ABCDEFG is presented in the text.}
\label{fig:FHN_path}
\end{figure}

Figure \ref{fig:FHN_path} shows a path in the stability (trace-versus-determinant) space for $0 \leq c \leq 3/2$. The path begins at $c = 0$ (A), where the fixed point is stable $(\tau < 0)$. As $c$ increases, it first reaches $c = 1/6$ (B) where the fixed point is at the critical point $x_{0} = -1$ of the $x$-nullcline. At $c = c_{s}(\epsilon) = 0.167167$ (C), the fixed point becomes marginally stable $(\tau = 0)$. A Hopf bifurcation yields a stable limit cycle for $c > c_{s}(\epsilon)$ as we go through $c = 3/4$ (D) until we return to marginal stability at $c = c_{u}(\epsilon) = 1.33283$ (E). As $c$ continues to increase, we reach $c = 4/3$ (F), when the fixed point is at the critical point $x_{0} = +1$ of the $x$-nullcline, and then ultimately we return to the starting point of the path at $c = 3/2$ (G). We note that at point D ($c = 3/4$), the trace $\tau$ reaches its highest (positive) value, which corresponds to the fastest firing rate.

\subsection{\label{sec:FHN_canard}Canard behavior of the FHN Solutions}

The singular canard perturbation equation \eqref{eq:canard_eq} for the FHN equations \eqref{eq:x_dot}-\eqref{eq:y_dot}   is
\begin{equation}
\epsilon \left(x + a - b\,\Phi\right) = \pd{\Phi}{x} \left( x - \frac{x^{3}}{3} + c \;-\; \Phi\right),
\end{equation}
where $\Phi(x,\epsilon) = \sum_{k=0}\epsilon^{k}\Phi_{k}(x)$ and $c(\epsilon) = \sum_{k=0}\epsilon^{k}c_{k}$, and we use $(a,b) = (3/5,4/5)$. Here, the partial derivatives evaluated at $\epsilon = 0$ are
\begin{equation}
\left. \begin{array}{rcl}
F_{y0} &=& -\,1 \\
F_{c0} &=& 1 \\
G_{y0} &=& -\,b \\ 
G_{c0} &=& 0
\end{array} \right\}.
\end{equation}
 At lowest order $(\epsilon = 0)$, we find $\Phi_{0}(x) = x - x^{3}/3 + c_{0}$, which has two critical points at $x_{c} = \pm 1$. 

At first order, we find
\begin{equation}
x + a - b\,\Phi_{0}(x) = \Phi_{0}^{\prime}(x)\left( c_{1} \;-\frac{}{} \Phi_{1}(x)\right),
\end{equation}
where the right side vanishes at the critical point $x = \pm 1$ of $\Phi_{0}(x)$. In order for $\Phi_{1}(x)$ to be regular at the critical points, we require the left side to also vanish at $x = \pm 1$. Hence, we find $12\,c_{0}^{\pm} - 9 = \pm 7$, which yields $c_{0}^{+} = c_{u}(0) = 4/3$ at $x_{c} = +1$ and $c_{0}^{-} = c_{s}(0) = 1/6$ at $x_{c} = -1$. By factoring both sides by $x \mp 1$, we find 
\[ H_{1}^{\pm}(x) \;=\; -\,\Psi_{0}^{\pm}(x)\,\left( c_{1} \;-\frac{}{} \Phi_{1}(x)\right), \]
where $H_{1}^{\pm}(x) = (4 x^{2} \pm 4 x + 7)/15$ and $\Psi_{0}^{\pm}(x) = (x \pm 1)$. Hence, the first-order solution is 
\begin{equation}
\Phi_{1}(x) \;=\; c_{1} \;+\; K_{1}(x), 
\label{eq:canard_1}
\end{equation}
where $K_{1}(x) = (4 x^{2} \pm 4 x + 7)/[15\,(x\pm 1)]$.

At the second order, we find
\[ -\,(4/5)\,\Phi_{1} \;=\; \Phi_{0}^{\prime}\,(c_{2} - \Phi_{2}) \;+\; \Phi_{1}^{\prime}\,(c_{1} - \Phi_{1}), \]
which can be expressed as
\begin{equation}
R_{2}(x) \;-\; (4/5)\,c_{1} \;=\; \Phi_{0}^{\prime}(x)\left( c_{2} \;-\frac{}{} \Phi_{2}(x)\right),
\end{equation}
where $R_{2}(x) = K_{1}(x)\,[K_{1}^{\prime}(x) - 4/5]$. Once again, since the right side vanishes at the critical point $x = \pm 1$ of $\Phi_{0}(x)$, we require that $c_{1}^{\pm} = R_{2}(\pm 1)/b = \mp 13/32$.

\begin{figure}
\epsfysize=3in
\epsfbox{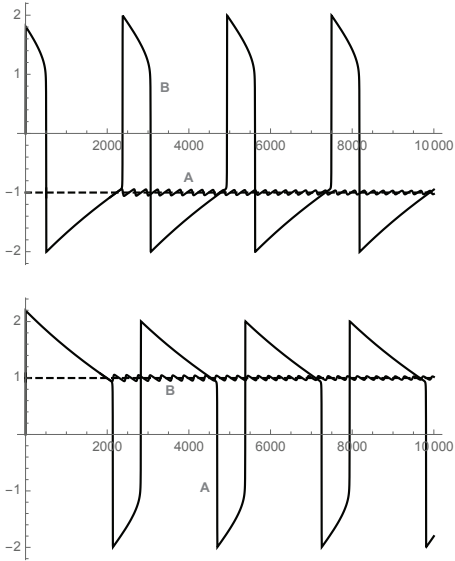}
\caption{Canard behavior in the FHN equations for $\epsilon = 0.001$ with initial conditions $(x,y) = (0,0)$. (top) Canard explosion $c = 0.16707\,(A) \rightarrow 0.16708\,(B)$ (bottom) Canard implosion $c = 1.33292\,(A) \rightarrow 1.33293\,(B)$.}
\label{fig:FHN_canard}
\end{figure}

Hence, when truncated at first order in $\epsilon$, the canard explosion and implosion occur at 
\begin{equation}
\left. \begin{array}{rcl}
c^{-}(\epsilon) &=& 1/6 \;+\; 13\,\epsilon/32 \;=\; 0.167073 \\
c^{+}(\epsilon) &=& 4/3 \;-\; 13\,\epsilon/32 \;=\; 1.33293
\end{array} \right\},
\end{equation}
respectively, where we used $\epsilon = 0.001$. These values agree very well with the numerical results shown in Fig.~\ref{fig:FHN_canard}. We note that the canard explosion occurs between points B ($c = 1/6$) and C ($c = c_{s}$) in Fig.~\ref{fig:FHN_path}, while the canard implosion occurs between points E ($c = c_{u}$) and F ($c = 4/3$), i.e., these canard events occur between marginal stability and the fixed point located at the critical points $\pm 1$ of the $x$-nullcline.

\subsection{\label{sec:FHN_period}Asymptotic limit-cycle period}

\begin{figure}
\epsfysize=1.5in
\epsfbox{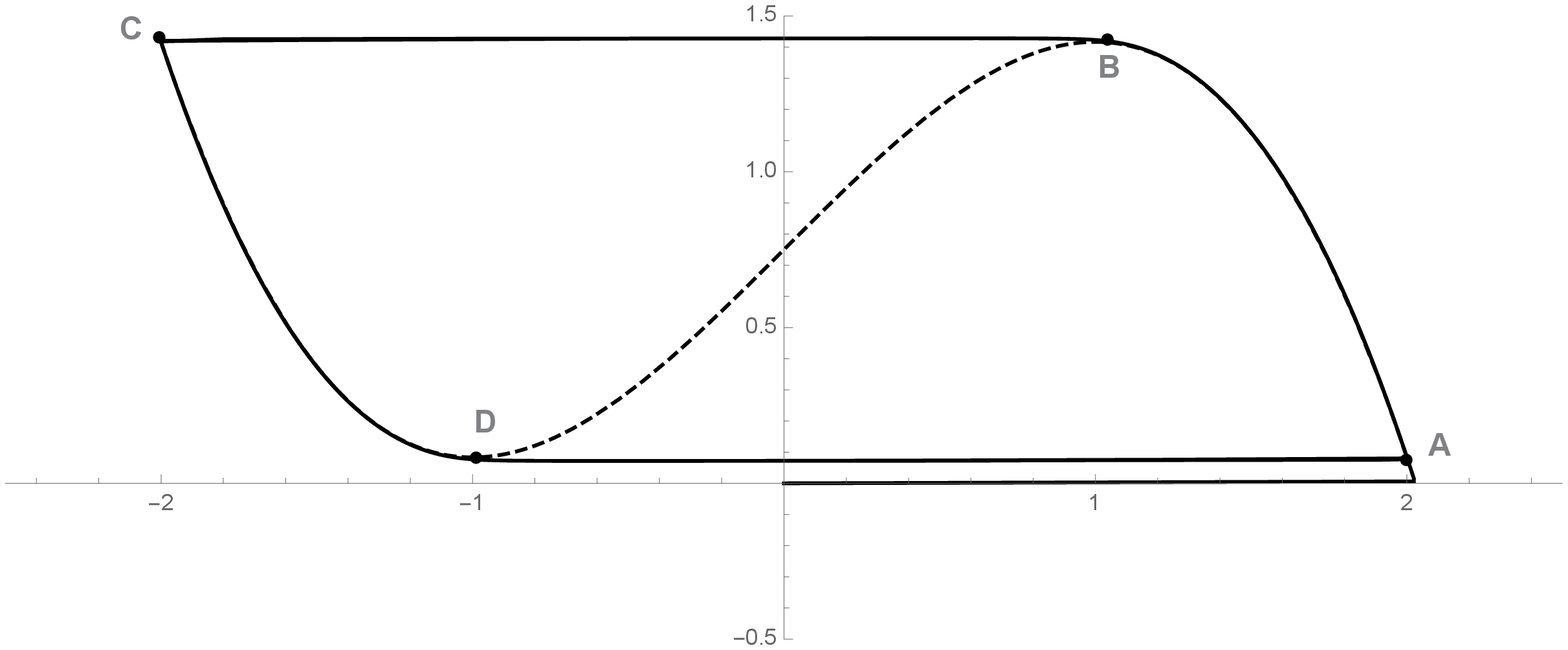}
\caption{Asymptotic limit cycle $ABCDA$ for the FHN equations, for $c = 3/4$ and $\epsilon = 0.001 \ll 1$. The slow segments $A\,(x = 2) \rightarrow B\,(x = 1)$ and $C\,(x = -2) \rightarrow D\,(x = -1)$ lie on the $x$-nullcline, while the transitions $B \rightarrow C$ and $D \rightarrow A$ are fast.}
\label{fig:FHN_phase}
\end{figure}

\begin{figure}
\epsfysize=1.8in
\epsfbox{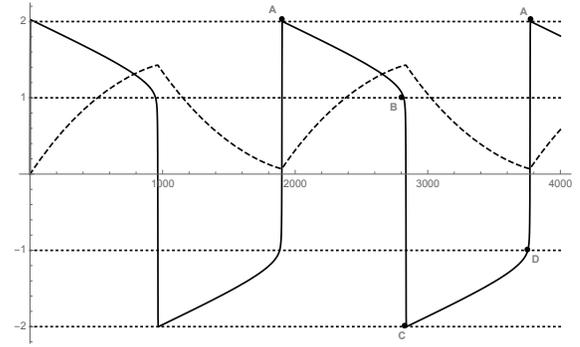}
\caption{FitzHugh-Nagumo solutions $x(t)$ (solid) and $y(t)$ (dashed) for $c = 3/4$ and $\epsilon = 0.001$.}
\label{fig:FHN_xy}
\end{figure}

Using the numerical solutions shown in Figs.~\ref{fig:FHN_phase}-\ref{fig:FHN_xy}, we see that the segments $A\,(x = 2) \rightarrow B\,(x = 1)$ and $C\,(x = -2) \rightarrow D\,(x = -1)$ on the $x$-nullcline occur on a much longer time scale than the fast transitions $B \rightarrow C$ and $D \rightarrow A$.

We now construct the asymptotic limit-cycle integral \eqref{eq:vdP_period} with the $x$-nullcline equation $y = x - x^{3}/3 + c$, which yields $\dot{y} = (1 - x^{2})\,\dot{x}$, and the $y$-equation $\dot{y} = \epsilon\,(x + a - b\,y)$ evaluated on the $x$-nullcline: $\dot{y} = \epsilon\,[x + a - b\,(x - x^{3}/3 + c)]$. We then obtain the infinitesimal equation
\begin{eqnarray}
\epsilon\;dt &=& \frac{(1 - x^{2})\,dx}{[b\,x^{3}/3 + (1-b)\,x - (bc - a)]} \nonumber \\
 &=& \frac{(3/b)\,(1 - x^{2})\,dx}{(x-x_{1})(x-x_{2})(x-x_{3})},
\end{eqnarray}
where $x_{1}(c) = x_{3}^{*}(c)$ and $x_{2}(c)$ are the roots defined in Eqs.~\eqref{eq:x1_c}-\eqref{eq:x3_c}.

\begin{figure}
\epsfysize=2in
\epsfbox{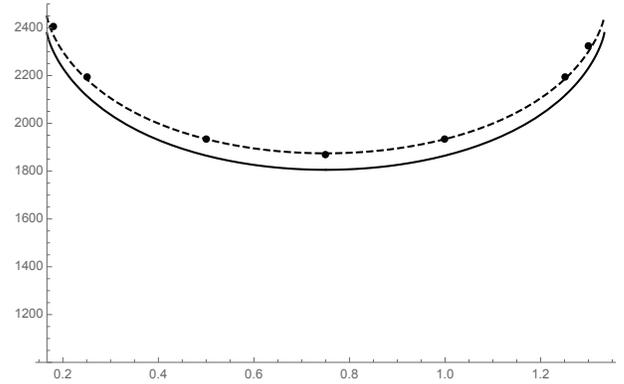}
\caption{Plots of the asymptotic FHN period $T_{\rm FHN}(c)$ (solid) and the corrected asymptotic FHN period $T_{\rm FHN}^{\alpha}(c,\epsilon)$ (dashed) versus $c$ for $\epsilon = 0.001$. Numerical periods (shown as dots) are all within 4\% above the function $T_{\rm FHN}(c)$, while they are within 1\% of the corrected period $T_{\rm FHN}^{\alpha}(c,\epsilon)$.}
\label{fig:FHN_period}
\end{figure}

The asymptotic limit-cycle period for the FHN equations is thus given by the integrals
\begin{eqnarray}
\epsilon\,T_{\rm FHN}(c) &=& \frac{3}{b}\int_{2}^{1}\frac{(1 - x^{2})\,dx}{(x-x_{1})(x-x_{2})(x-x_{3})} 
\label{eq:FHN_period_def} \\
 &&+\; \frac{3}{b}\int_{-2}^{-1}\frac{(1 - x^{2})\,dx}{(x-x_{1})(x-x_{2})(x-x_{3})}. \nonumber
\end{eqnarray}
We now introduce the partial-fraction decomposition
\[ \frac{(1 - x^{2})}{(x-x_{1})(x-x_{2})(x-x_{3})} = \frac{p_{1}}{x - x_{1}} + \frac{p_{2}}{x - x_{2}} + \frac{p_{3}}{x - x_{3}}, \]
 with the coefficients
 \begin{eqnarray}
 p_{1}(c) & = & -\,\frac{(x_{2} - x_{3})}{\Delta}\;\left(1 \;-\frac{}{} x_{1}^{2} \right), \\
 p_{2}(c) & = & -\,\frac{(x_{3} - x_{1})}{\Delta}\;\left(1 \;-\frac{}{} x_{2}^{2} \right), \\
 p_{3}(c) & = & -\,\frac{(x_{1} - x_{2})}{\Delta}\;\left(1 \;-\frac{}{} x_{3}^{2} \right),
 \end{eqnarray}
 where $\Delta = (x_{1} - x_{2})\,(x_{2} - x_{3})\,(x_{3} - x_{1})$ and we used $x_{1} + x_{2} + x_{3} = 0$. Hence, Eq.~\eqref{eq:FHN_period_def}  can be written as
\begin{equation} 
\epsilon\,T_{\rm FHN}(c) \;=\; -\;\frac{3}{b}\sum_{k=1}^{3}p_{k}\;\ln\left( \frac{4 - x_{k}^{2}}{1 - x_{k}^{2}}\right).
\label{eq:FHN_period}
\end{equation}
If we add the same nontrivial Van der Pol correction $3\alpha\,\epsilon^{2/3}$ [see Eq.~\eqref{eq:VdP_period_alpha}] to the asymptotic FHN period \eqref{eq:FHN_period}, we can define the corrected period
\begin{equation} 
T^{\alpha}_{\rm FHN}(c,\epsilon) \;\equiv\; T_{\rm FHN}(c) \;+\; 3\,\alpha/\epsilon^{1/3}.
\label{eq:FHN_period_alpha}
\end{equation}
Figure \ref{fig:FHN_period} shows that the exact numerical periods (shown as dots) are all within 4\% above the period $T_{\rm FHN}(c)$, while they are within 1\% of the $\alpha$-corrected period $T_{\rm FHN}^{\alpha}(c,\epsilon)$. 

\section{Summary}

In the present paper, we have shown how the asymptotic limit-cycle properties of the FHN equations \eqref{eq:x_FHN}-\eqref{eq:y_FHN} can be accurately calculated. Indeed, we have shown in Sec.~\ref{sec:FHN_canard} how the singular perturbation theory of Fenichel \cite{Fenichel_1979} can be used to accurately predict the appearance (canard explosion) and disappearance (canard implosion) of large-amplitude relaxation oscillations (see Fig.~\ref{fig:FHN_canard}) in the FHN equations \eqref{eq:x_FHN}-\eqref{eq:y_FHN}. In addition, once large-amplitude relaxation oscillations are excited, the period of these oscillations can be accurately calculated in Eq.~\eqref{eq:FHN_period}, where explicit formulas for the cubic roots \eqref{eq:x1_c}-\eqref{eq:x3_c} of the polynomial \eqref{eq:FHN_cubic}. The accuracy of Eq.~\eqref{eq:FHN_period_alpha} is clearly demonstrated in Fig.~\ref{fig:FHN_period} when the nontrivial Van der Pol correction $3\alpha\,\epsilon^{2/3}$ is added to the FHN period \eqref{eq:FHN_period}.

\end{document}